\documentclass{PoS}
\usepackage{macros,macros_static}
\usepackage{amstext,amsfonts,amsbsy,amsmath,amssymb,amscd,epsf,mathrsfs,dsfont}
\usepackage{rotating}
\usepackage{multicol}
\usepackage{pstricks}
\usepackage{pst-node}
\usepackage{latexsym}
\usepackage{epsfig}
\usepackage{ulem} 
\def\Oa{O($a$) }
\def\mpi2{m_\pi^2}
\def\cQ{{\cal Q}}

\def\cO{{\cal O}}

\def\cB{{\cal B}}

\def\half{\frac{1}{2}}

\newcommand{\cR}{{\cal R}}
\newcommand{\cS}{{\cal S}}

\newcommand{\pcd}{\Delta}

\newcommand{\<}{\langle}

\newcommand{\msbar}{{\overline{\rm MS}}}
\renewcommand{\>}{\rangle}

\newcommand{\bflr}{\begin{flushright}}
\newcommand{\eflr}{\end{flushright}}
\newcommand{\bfll}{\begin{flushleft}}
\newcommand{\efll}{\end{flushleft}}
\newcommand{\ben}{\begin{enumerate}}
\newcommand{\een}{\end{enumerate}}
\newcommand{\beq}{\begin{equation}}
\newcommand{\eeq}{\end{equation}}
\newcommand{\beqn}{\begin{eqnarray}}
\newcommand{\eeqn}{\end{eqnarray}}
\newcommand{\nn}{\nonumber}

\newcommand{\psih}{\psi_{\scriptscriptstyle\mathbf{h}}}
\newcommand{\psihb}{\psi_{\scriptscriptstyle\mathbf{\bar h}}}
\newcommand{\bpsih}{\bar\psi_{\scriptscriptstyle\mathbf{h}}}
\newcommand{\bpsihb}{\bar\psi_{\scriptscriptstyle\mathbf{\bar h}}}
\newcommand{\ci}[1]{\boldsymbol{#1}}

\title{Renormalization of HQET $\Delta B=2$ operators:\\O($a$) improvement and $1/m$ matching with QCD}

\ShortTitle{Renormalization of HQET $\Delta B=2$ operators}

\author{\speaker{Mauro Papinutto}\\
         Dip. di Fisica, "Sapienza" Universit\`a di Roma and INFN, Sezione di Roma, P.le A. Moro 2, I-00185 Roma, Italy\\
        E-mail: \email{mauro.papinutto@roma1.infn.it}}

\author{Gregorio Herdoiza\\
            PRISMA Cluster of Excellence, Institut f\"ur Kernphysik, Johannes Gutenberg-Universit\"at,
D-55099 Mainz, Germany\\
	E-mail: \email{herdoiza@kph.uni-mainz.de}}

\author{Carlos Pena\\
             Instituto de F\'\i sica Te\'orica UAM/CSIC and Departamento de F\'\i sica Te\'orica, Universidad
Aut\'onoma de Madrid, Cantoblanco E-28049 Madrid, Spain\\
         E-mail: \email{carlos.pena@uam.es}}

\author{Anastassios Vladikas\\
        INFN, Sezione di Tor Vergata, c/o Dipartimento di Fisica, Universit\`a di Roma Tor Vergata,\\ Via della Ricerca Scientifica 1, I-00133 Rome, Italy\\
        E-mail: \email{tassos.vladikas@roma2.infn.it}}

\abstract{We determine a basis of dimension-7 operators which arise at O($a$) in the Symanzik expansion of the $\Delta B=2$ operators with static heavy quarks. We consider both Wilson-like and Ginsparg-Wilson light quarks. Exact chiral symmetry reduces the number of these O($a$) counterterms by a factor of two. Only a subset of these operators has previously appeared in the literature. We then extend the analysis to the O($1/m$) operators contributing beyond the static approximation.}

\FullConference{31st International Symposium on Lattice Field Theory LATTICE 2013\\
		 July 29 -- August 3, 2013\\
		 Mainz, Germany}

\begin{document}

\section{Introduction}
The experimental accuracy in $B$-physics measurements is being remarkably
improved, mainly thanks to the performance of LHCb. 
A reliable computation of the $B$-parameters $B_{\rm B_d}$ and $B_{\rm B_s}$,
as well as other $B$-parameters appearing in extensions beyond the Standard Model 
(BSM) is essential when constraining the CKM and BSM parameters, by combining experimental 
results and theory predictions.

The $b$-quark is too heavy to be simulated directly on presently
attainable lattice spacings. In this study we use Heavy Quark Effective 
Theory (HQET), an expansion in inverse powers
of the heavy quark mass $m$, which is well suited to a non-perturbative
treatment of  the $b$-quark with controlled systematic uncertainties.
Since HQET is an effective theory, it can be renormalized at fixed
order in $1/m$, by determining a finite number of parameters. 
At this order, the lattice-regularized theory can be
renormalized (i.e. it has a well defined continuum limit) and the parameters
can be determined by matching with QCD in a small volume
(see~\cite{Heitger:2003nj} and notation therein).

In the present study we apply HQET to the computation of the 
$B$-parameters of the neutral $B$-meson system. 
In order to reduce systematic uncertainties at the level of few percent,
we investigate the matching of HQET to QCD at O($\minv$). Defining the static
b-quark fields as
$H\equiv \heavy+\aheavy$ and  $\bar H\equiv\heavyb+\aheavyb$, the HQET action
and a generic (multiplicatively renormalizable) composite operator read:
\beqn
\label{eq:HQETaction}
  \Shqet &=& a^4 \sum_x \{\bar H \gamma_0 D_0 H - 
           \omegaspin \underbrace{\bar H \vecsigma\cdot \vecB H}_{\equiv \op{spin}} -
           \ \omegakin \underbrace{\bar H {\frac12 {\bf \nabla}^2 } H}_{\equiv
             \op{kin}}
           \,+\  {\rm O}(1/m^2) \},\\
{\cal O}^{\rm QCD} &=& Z^{\rm HQET}_{\cal O} \{ {\cal O}^{\rm stat}+ c^{i}_{\cal O} {\cal O}_i^{1/m} + {\rm O}(1/m^2)\},
\eeqn
\noindent with $\{\omegakin,\ \omegaspin,\ c^{i}_{\cal O}\} = {\rm O}(\minv)$ and
$Z^{\rm HQET}_{\cal O}$ to be determined in the matching procedure.

As pointed out in~\cite{Heitger:2003nj}, the path integral can be written as
an ensemble average with respect to the static action if, in the spirit of the
expansion, O($\minv$) terms are treated as operator insertions:
\beqn
\label{eq:O_1oM}
   \langle {\cal{O}} \rangle^{\rm QCD} &=& Z^{\rm HQET}_{\cal O}\{\< {\cal O}^{\rm stat}  \>_{\rm stat} + \omegaspin \,a^4\sum_x \< {\cal O}^{\rm stat} {\cal O}_{\rm spin}(x) \>_{\rm stat} +\nn\\
&+& \omegakin \, a^4\sum_x  \<{\cal O}^{\rm stat} {\cal O}_{\rm kin}(x) \>_{\rm stat} + c^{i}_{\cal O} \< {\cal O}_i^{1/m}\>_{\rm stat}+  {\rm O}(1/m^2)\}.
\eeqn
The $N_{\cal O}^{\rm HQET}$ bare couplings $\omegakin$, $\omegaspin$,
  $Z^{\rm HQET}_{\cal O}$, $c^{i}_{\cal O}$, $\dots$ of HQET at O($\minv$)
can be determined by imposing matching conditions between suitably chosen operator correlation functions $\Phi^{\rm QCD}_k(m,L)$  and $\Phi^{\rm HQET}_k(m,L)$, defined in small volume QCD and HQET respectively: 
\beqn
\label{eq:matching}
\Phi^{\rm QCD}_k(m,L) = \Phi^{\rm HQET}_k(m,L)+{\rm O}\left(\frac{1}{m^2}\right)\,,\quad k=1,2,\ldots,N_{\cal O}^{\rm HQET}
\eeqn
where the linear size $L$ used in practice is $L\approx 0.4\,\fm$ (the  parameters of the 
QCD and HQET Lagrangians are volume independent). The l.h.s. is a quantity defined in the 
continuum, and computed in lattice QCD for lattice spacings $a$, chosen to be small enough 
so as to fulfill the condition $m_{\rm b} a \ll 1$. This ensures that in $\Phi^{\rm QCD}_k(m,L)$
the $b$-quark can be simulated relativistically.
The r.h.s. is defined at finite lattice spacing $a$. Note that in HQET the relativistic $b$-quark is no more 
present and thus constraints on the size of $a$ are absent. The connection to large volume 
where the B-meson fits comfortably and physical observables (e.g. $B_{\rm B_d}$, $\Bbs$, $F_{\rm B_d}$, $\Fbs$) can be computed, is done recursively in HQET using a step scaling function procedure in the Schr\"odinger Functional (SF) scheme~\cite{Heitger:2003nj}.

\section{$\Delta B=2$ four-fermion operators in HQET}
\label{sec:dim6}

The $B_{\rm B_q}$ parameters which enter SM predictions are defined to be
\beqn
B_{\rm B_q}(\mu)= \frac{\langle {\rm {\bar B}_q}^0|
{\cal O}_{LL}^{\Delta B=2}(\mu)|{\rm B_q}^0\rangle}{\frac{8}{3}F_{\rm B_q}^2 m_{\rm B_q}^2}\qquad \mathrm{with}\qquad 
{\cal O}_{LL}^{\Delta B=2}=\bar{b}\gamma_{\mu}(1-\gamma_5)q \;\bar{b}\gamma_{\mu}(1-\gamma_5)q
\eeqn 
and similarly those which appear in models of new physics (the latter are 
related to matrix elements of other $\Delta B = 2$ four-fermion operators, see~\cite{Becirevic:2001xt}).
 
In HQET, a complete basis of 6 dimensional $\Delta H = 2$ operators  
(with two relativistic light quarks $\psi$ of the same flavor) is:
\begin{gather}
\begin{array}{cccccc}
Q_1 &\,\,=\,\,& \cO_{\rm VV+AA}\, \quad &\cQ_1 &\,\,=\,\,& \cO_{\rm VA+AV}\,\\[1.0ex]
Q_2 &\,\,=\,\,& \cO_{\rm SS+PP}\, \quad &\cQ_2 &\,\,=\,\,& \cO_{\rm SP+PS}\,\\[1.0ex]
Q_3 &\,\,=\,\,& \cO_{\rm VV-AA}\, \quad &\cQ_3 &\,\,=\,\,& \cO_{\rm VA-AV}\,\\[1.0ex]
Q_4 &\,\,=\,\,& \cO_{\rm SS-PP}\, \quad &\cQ_4 &\,\,=\,\,& \cO_{\rm SP-PS}\,
\end{array}\\[1.5ex]
\mathrm{with}\quad \cO_{\Gamma_1\Gamma_2\pm \Gamma_3\Gamma_4} = \left[
(\bpsih\Gamma_1\psi)(\bpsihb\Gamma_2\psi)\,\pm\,
(\bpsih\Gamma_3\psi)(\bpsihb\Gamma_4\psi) \right]
\end{gather} 
where the $Q_i$ are parity even (PE) while $\cQ_i$ are parity odd (PO). 

The static heavy quark action is invariant under spatial rotations H(3), rather than the usual rotations H(4), and under the heavy quark spin symmetry (HQSS):
\begin{gather}
\psih(x)~\to~e^{i\alpha_k\Sigma_k}\,\psih(x)\equiv\cS(\alpha)\psih(x)~~~~~~~
\bpsih(x)~\to~\bpsih(x)\,e^{-i\alpha_k\Sigma_k}\equiv\bpsih(x)\cS(\alpha)^\dag
\end{gather}
where $\Sigma_k=\half\epsilon_{ijk}\sigma_{ij}=\frac{i}{2}\epsilon_{ijk}\gamma_i\gamma_j$,
and $\psihb$ and $\bpsihb$ transform analogously with $\beta_k$ independent of $\alpha_k$.
By using HQSS one can prove~\cite{Becirevic:2003hd,Palombi:2006pu} that with Wilson-like 
fermions~\footnote{i.e. standard or improved Wilson fermions or twisted mass fermions} the new basis 
\begin{gather}
\label{6dim_basis}
\begin{array}{cclccl}
Q_1' &\,\,=\,\,& Q_1 \quad &\cQ_1' &\,\,=\,\,& \cQ_1\\[1.0ex]
Q_2' &\,\,=\,\,&  Q_1+4Q_2\quad &\cQ_2' &\,\,=\,\,& \cQ_1+4\cQ_2\\[1.0ex]
Q_3' &\,\,=\,\,& Q_3+2 Q_4 \quad &\cQ_3' &\,\,=\,\,& \cQ_3+2\cQ_4\\[1.0ex]
Q_4' &\,\,=\,\,& Q_3-2 Q_4 \quad &\cQ_4' &\,\,=\,\,&\cQ_3-2\cQ_4 
\end{array}
\end{gather} 
simplifies the mixing pattern. For the PE sector we have: 
\begin{gather}
\label{diagmixing}
\left(\begin{array}{c}
	  Q_1' \\
	  Q_2' \\
	  Q_3' \\
	  Q_4'
	\end{array}\right)^{\rm R}
	=
	\left(\begin{array}{cccc}
	  Z_{1} & 0           & 0           & 0          \\
	  0          & Z_{2}  & 0           & 0          \\
	  0          & 0           & Z_{3}  & 0 \\
	  0          & 0           & 0           & Z_{4}
	\end{array}\right)
	{\left[ \mathds{1} + 
	\left(\begin{array}{cccc}
	  0          & 0          & \pcd_{1}   & 0 \\
	  0          & 0          & 0          & \pcd_{2} \\
	  \pcd_{3}   & 0          & 0          & 0 \\
	  0          & \pcd_{4}   & 0          & 0
	\end{array}\right)
	\right]}
	\left(\begin{array}{c}
	  Q_1' \\
	  Q_2' \\
	  Q_3' \\
	  Q_4'
	\end{array}\right)
\end{gather}
while, by using time reversal, one can show that the PO sector renormalizes multiplicatively~\cite{Palombi:2006pu}.

With Ginsparg-Wilson fermions, both the PO and PE sectors renormalize
multiplicatively, the discrete axial transformation ${\cal R}_5$: $\psi
\rightarrow \gamma_5 \psi$, $\bar\psi \rightarrow -\bar\psi\gamma_5$ beeing an
exact symmetry which implies:
\begin{gather}
\begin{array}{cccccccccccc}
Q_1' &\,\,\rightarrow\,\,& Q_1' \quad & Q_2' &\,\,\rightarrow\,\,&Q_2'   \quad &
Q_3' &\,\,\rightarrow\,\,& -Q_3'  \quad &Q_4' &\,\,\rightarrow\,\,& -Q_4'\\
\cQ_1' &\,\,\rightarrow\,\,& \cQ_1' \quad &\cQ_2' &\,\,\rightarrow\,\,& \cQ_2' \quad&
\cQ_3' &\,\,\rightarrow\,\,& -\cQ_3' \quad & \cQ_4' &\,\,\rightarrow\,\,& -\cQ_4'
\end{array}
\end{gather}

Here we will investigate the strategy to compute matrix elements of $\Delta B = 2$ four-fermion operators using the example of ${\cal O}_{LL}^{\Delta B=2}$. This operator is even under the axial transformation $\cR_5$. Consequently, in matching QCD to HQET, we need to consider only $(Q_1')^{\rm R}$ and $(Q_2')^{\rm R}$. However, when working with Wilson-like 
fermions, these two operators mix with $Q_3'$ and $Q_4'$ under renormalization.
As explained in~\cite{Palombi:2006pu}, a way out of this spurious mixing is the use of twisted mass QCD at maximal twist ($\alpha=\pi/2$), with $\psi$ a component of an isospin doublet of light quarks. 
In this way the operators $Q_1'$ and $Q_2'$ are mapped into $\cQ_1'$ and $\cQ_2'$ and the matrix elements can be computed according to
\beqn
\label{eq:tmQCD}
\<\bar B\, Q_1'^{\rm R}\, B\>^{\rm Wilson}\ \ =\ -i \<\bar \cB\,\cQ_1'^{\rm R}\,\cB\>^{\rm tmQCD}\qquad\quad \<\bar B\, Q_2'^{\rm R}\, B\>^{\rm Wilson}\ \ =\ -i \<\bar \cB\, \cQ_2'^{\rm R}\, \cB\>^{\rm tmQCD}
\eeqn 
where $B$, $\bar B$, $\cB$ and $\bar \cB$ are suitable source and sink operators for the $B$ mesons and $\cQ_1'$ and $\cQ_2'$ renormalize multiplicatively and have the correct chiral properties. In~\cite{Palombi:2007dr,Dimopoulos:2007ht} the corresponding renormalization constants and renormalization group running have been computed with $N_f=0$ and $N_f=2$ dynamical flavors. We leave for the moment the question open and we investigate the structure of \Oa and O($\minv$) terms. 

\section{O($a$) improvement}

We first consider the improvement terms in the massless limit. 
Using the equations of motion, Fierz identities and the 
properties of heavy quark fields (including the local flavour transformation
$\psih(x)\rightarrow e^{i\eta({\bf x})}\psih(x)$, $\bpsih(x)\rightarrow
\bpsih(x) e^{-i\eta({\bf x})}$ which is a symmetry of the 
static action) we obtain a basis of independent dimension-7 
PE operators:
\begin{gather}
\begin{array}{ccccc}
\gamma_j \otimes D_j &\quad & \gamma_j \gamma_5  \otimes \gamma_5 D_j &\quad&  \epsilon_{ijk}\  [\gamma_i  \otimes \gamma_j \gamma_5 D_k]\\
I  \otimes \gamma_j D_j  &\quad& \gamma_5 \otimes \gamma_j \gamma_5 D_j&\quad&
\epsilon_{ijk}\ [\gamma_i\gamma_5 \otimes \gamma_j D_k]\\
D_j\otimes \gamma_j &\quad & \gamma_5 D_j  \otimes \gamma_j\gamma_5 &\quad&  \epsilon_{ijk}\  [\gamma_i \gamma_5 D_j\otimes \gamma_k ]\\
\gamma_j  D_j\otimes I&\quad& \gamma_j\gamma_5 D_j \otimes \gamma_5&\quad&
\epsilon_{ijk}\ [\gamma_i D_j\otimes \gamma_k\gamma_5]
\end{array}
\end{gather}
where we have used the concise notation
\beqn
\Gamma_i \otimes \Gamma_j D_k\ \equiv\ (\bpsih\Gamma_i\psi)(\bpsihb\Gamma_j D_k\psi), \qquad \Gamma_i D_j\otimes \Gamma_k\ \equiv\ (\bpsih\Gamma_i D_j\psi)(\bpsihb\Gamma_k\psi)
\eeqn
The (spatial) covariant derivative $D_i$ acts on the light quark
field only, since local flavor conservation in the static action forbids terms
containing spatial derivatives of the heavy quark fields.  
\noindent We can now pass from these operators to a new set of 12 operators through the identities: \begin{eqnarray}
\gamma_i \otimes \gamma_i (\ci{\gamma}\cdot\mathbf{D})&= &
\gamma_i \otimes  D_i - \epsilon_{ijk} [ \gamma_i \otimes \gamma_j \gamma_5 D_k ] \nn\\
\gamma_i \gamma_5 \otimes \gamma_i \gamma_5 (\ci{\gamma}\cdot\mathbf{D})&= & -\gamma_i\gamma_5 \otimes \gamma_5 D_i + \epsilon_{ijk} [ \gamma_i \gamma_5 \otimes \gamma_j D_k ]\nn \\
\gamma_0 \otimes \gamma_0 (\ci{\gamma}\cdot\mathbf{D}) &=& - 1  \otimes \gamma_i D_i \nn\\
\gamma_0 \gamma_5 \otimes \gamma_0\gamma_5 (\ci{\gamma}\cdot\mathbf{D})&= & \gamma_5  \otimes \gamma_i \gamma_5 D_i \\
\gamma_i \otimes  (\ci{\gamma}\cdot\mathbf{D}) \gamma_i &= &
\gamma_i \otimes D_i + \epsilon_{ijk} [ \gamma_i \otimes \gamma_j \gamma_5 D_k ]\nn\\
\gamma_i \gamma_5 \otimes (\ci{\gamma}\cdot\mathbf{D}) \gamma_i \gamma_5 &=  &\gamma_i\gamma_5 \otimes \gamma_5 D_i  + \epsilon_{ijk} [ \gamma_i \gamma_5 \otimes \gamma_j D_k] \nn
\end{eqnarray}
plus the analogous ones with the scalar product $(\ci{\gamma}\cdot\mathbf{D})$ in the first bilinear.

We can then combine these twelve operators to obtain six with the same time reversal properties of $Q_1'$ and $Q_2'$:
\begin{gather}
\begin{array}{ccc}
\delta Q_1&=&   \gamma_i \otimes \gamma_i (\ci{\gamma}\cdot\mathbf{D})+\gamma_i (\ci{\gamma}\cdot\mathbf{D})\otimes\gamma_i\\
\delta Q_2&=&\gamma_i \gamma_5\otimes \gamma_i \gamma_5(\ci{\gamma}\cdot\mathbf{D}) +\gamma_i \gamma_5 (\ci{\gamma}\cdot\mathbf{D}) \otimes\gamma_i \gamma_5 \\
\delta Q_3&=&\gamma_0 \otimes \gamma_0 (\ci{\gamma}\cdot\mathbf{D})+\gamma_0 (\ci{\gamma}\cdot\mathbf{D})\otimes \gamma_0 \\
\delta Q_4&=&\gamma_0 \gamma_5 \otimes \gamma_0 \gamma_5(\ci{\gamma}\cdot\mathbf{D})+\gamma_0 \gamma_5 (\ci{\gamma}\cdot\mathbf{D})\otimes \gamma_0 \gamma_5\\
\delta Q_5&=&\gamma_i \otimes (\ci{\gamma}\cdot\mathbf{D}) \gamma_i +(\ci{\gamma}\cdot\mathbf{D})\gamma_i \otimes  \gamma_i \\
\delta Q_6&=&\gamma_i \gamma_5\otimes (\ci{\gamma}\cdot\mathbf{D})\gamma_i\gamma_5+(\ci{\gamma}\cdot\mathbf{D})\gamma_i \gamma_5\otimes  \gamma_i \gamma_5
\end{array}
\end{gather}

We now want to further constrain the form of the counterterms by using the remaining symmetries of the HQET action (namely H(3) rotations and HQSS). We will briefly sketch here the method used to constrain the renormalization pattern of dimension-6 operators, presented in Appendix~A of~\cite{Palombi:2006pu} and extend it to include also the O($a$) counterterms.  

Let us consider the vector of operators $\vec Q$ and the vector of O($a$) 
counterterms $\delta\vec Q$ with $n$ and $m$ components respectively. The components of
$\vec Q$ can (in principle) mix with scale-dependent coefficients, as well as with
the counterterms $\delta\vec Q$ with scale-independent coefficients.
The $n$-vector of (improved) renormalized operators $\vec Q_{\rm R}$ will have the form
\begin{gather}
\label{eq:vec_renorm}
\vec Q_{\rm R} = Z(\vec Q + c\,\delta\vec Q)\,,
\end{gather}
where $Z$ is an $n \times n$ renormalization matrix and $c$ is an $n \times m$
mixing matrix. Let us now consider a transformation $S$, realised
at the level of operator vectors as a linear transformation with matrices
$\Phi_{S}$ and $\tilde\Phi_{S}$
($n \times n$ and $m \times m$, respectively), viz.
\begin{gather}
\vec Q~~\to~~\Phi_{S}\vec Q\,,~~~~~~~~~~~
\delta\vec Q~~\to~~\tilde\Phi_{S}\delta\vec Q\,.
\end{gather}
If $S$ is a symmetry transformation, a natural way of enforcing the symmetry
at the level of the renormalized theory 
%(i.e. of ensuring that the relevant Ward-Takahashi identities have the correct form) 
is to require that renormalized operators
transform in the same way as the corresponding bare operators. This can be stated as
\begin{gather}
\Phi_{S}\vec Q_{\rm R} = Z(\Phi_{S}\vec Q + c\,\tilde\Phi_{S}\delta\vec Q)\,.
\end{gather}
Consistency with Eq.~(\ref{eq:vec_renorm}) then requires
\begin{gather}
\label{eq:symm_cond}
Z=\Phi_{S}\, Z\, \Phi_{S}^{-1}\,,~~~~~~~~~~~~~
c=\Phi_{S}\, c\, \tilde\Phi_{S}^{-1}\,,
\end{gather}
where the two separate conditions on $Z$ and $c$ ensure that $c$ is kept
scale-independent.
In practice, one will choose a sufficiently large set of symmetry transformations
$\{S_i\}$, and impose Eqs.~(\ref{eq:symm_cond}) on generic $Z$ and $c$ matrices. If the
matrices $\Phi_{S_i}$ have non-trivial blocks, the resulting constraints will reduce the number of independent entries of the matrices $Z$ and $c$. One can then look
for a basis which maximises the number of zeroes in the matrices, so that the
final result is as simple as possible.
By choosing $\{S_i\}=\{{\cal S}_1,\cR_1,\cR_3\}$ (with $\cS_1$ a HQSS rotation with $\vec \alpha=-\vec \beta = \{\pi/2,0,0\}$ and $\cR_{1,3}$ rotations by $\pi/2$ around axes $\hat 1$ and $\hat 3$ respectively) one can show that the counterterm structure of the first two operators $Q_1'$ and $Q_2'$ is the following:
\begin{gather}
\label{eq:Oa_c1}
Q_1'\qquad\qquad\left\{ 
\begin{array}{ccl}
\delta Q'_{1}&=& \delta Q_1-\delta Q_2-\delta Q_3+\delta Q_4\\
\delta Q'_{2}&=& \delta Q_1+\delta Q_2+\delta Q_3+\delta Q_4
\end{array}\right .
\end{gather}
\begin{gather}
\label{eq:Oa_c2}
Q_2'\qquad\qquad\left\{ 
\begin{array}{ccl}
\delta Q'_{3}&=& \delta Q_1-\delta Q_2+3\delta Q_3-3\delta Q_4\\
\delta Q'_{4}&=& \delta Q_3-\delta Q_4+\delta Q_5+\delta Q_6\\
\delta Q'_{5}&=& \delta Q_1+\delta Q_2-3\delta Q_3-3\delta Q_4\\
\delta Q'_{6}&=& \delta Q_3+\delta Q_4-\delta Q_5+\delta Q_6
\end{array}\right .
\end{gather}
with $c$ a $2\times6$ matrix where only $c_{11}$, $c_{12}$, $c_{23}$, $c_{24}$, $c_{25}$, $c_{26}$
are non-zero.

Counterterms $\delta Q'_{1}$, $\delta Q'_{3}$ and $\delta Q'_{4}$ are $\cR_5$-even whereas $\delta Q'_{2}$, $\delta Q'_{5}$ and $\delta Q'_{6}$ are $\cR_5$-odd.  
Hence, for lattice formulations with exact chiral symmetry, $\cR_5$ implies that $\delta Q'_{2}$, $\delta Q'_{5}$ and $\delta Q'_{6}$ do not contribute and one is left with 
the sinlge counterterm $\delta Q'_{1}$ for $Q_1'$, and the two counterterms $\delta Q'_{3}$, $\delta Q'_{4}$ for $Q_2'$.
 
In~\cite{Ishikawa:2011dd} the one-loop perturbative matching including O($a$) improvement is computed for domain-wall fermions. We explicitly checked that our results agree with theirs for the operators $\delta Q'_1$ and $\delta Q'_3$. However, the operator $\delta Q'_{4}$ is missing in~\cite{Ishikawa:2011dd}. For Wilson fermions,  our results are in agreement with theirs, except for the two operators $\delta Q'_4$ and $\delta Q'_{6}$ which are again missing, cf. Appendix B in~\cite{Ishikawa:2011dd}. Naturally the question arises: do these operators appear in perturbation theory only at two loops?

Imposing HQSS, we obtain the following O($a$) counterterms which are proportional to the light quark mass $m_\ell$:
\begin{gather}
b_{11}\, m_\ell\, Q_1' + b_{13}\, m_\ell\, Q_3'  \qquad\textrm{for}\qquad Q_1'\nn\\
b_{22}\, m_\ell\, Q_2' + b_{24}\, m_\ell\, Q_4'  \qquad\textrm{for}\qquad Q_2'
\label{eq:Oa_b}
\end{gather}
In the case of Ginsparg-Wilson fermions, the spurionic symmetry $(m_\ell \rightarrow -m_\ell)\times \cR_5$ 
implies that $b_{11}=b_{22}=0$. Our conclusions agree with the results of Ref.~ \cite{Ishikawa:2011dd}.

\vspace*{-0.1cm}
\section{O($1/m$) terms and conclusions}

The $1/m$ terms in the expansion Eq.~(\ref{eq:O_1oM}) are found by considering dimension-7 operators with the common symmetries of ${\cal O}^{\rm stat}$ and of the HQET action. These are $H(3)$ cubic invariance, parity, time reversal and flavor. HQSS and local flavor conservation are not symmetries of ${\cal O}_{\rm spin}$, $ {\cal O}_{\rm kin}$ respectively.
The $1/m$ terms are obtained by taking the O($a$) counterterms in
Eq.~(\ref{eq:Oa_c1}),~(\ref{eq:Oa_c2})
with $\mathbf{D}$ replaced either by $(\mathbf{D} -
\overleftarrow{\mathbf{D}})$ or by $(\mathbf{D} +
\overleftarrow{\mathbf{D}})$ plus the four operators coming from Eq.~(\ref{eq:Oa_b}).
The operators in Eq.~(\ref{eq:Oa_c1}),~(\ref{eq:Oa_c2}) are thus doubled to a total amount 
of twelve operators. 

Again, with Ginsparg-Wilson fermions only half of the previous dimension-7
operators would contribute, i.e. only those with the correct 
$(m_\ell \rightarrow - m_\ell)\times\cR_5$ symmetry.

In order to compute the matrix element of ${\cal O}_{LL}^{\Delta B=2}$ from
HQET at O($1/m$) we have thus to consider the four dimension-6 operators $Q_i'$ in
Eq.~(\ref{6dim_basis}) plus the twelve dimension-7 operators originating from Eq.~(\ref{eq:Oa_c1}),~(\ref{eq:Oa_c2}) while the four dimension-7 operators in 
Eq.~(\ref{eq:Oa_b}) will amount to as mass-dependent redefinition of the matching 
coefficients of the dimension-6 operators. This will require 16 matching conditions through which, in the language of Eq.~(\ref{eq:O_1oM}),~(\ref{eq:matching}), the 4 $Z^i_{\cO_{LL}}$ and the 12 $c^i_{\cO_{LL}}$ coefficients have to be determined. Moreover one has to determine 
$m_{\rm bare}$, $\omegakin$, $\omegaspin$ which requires 3 further conditions\footnote{We did not include explicitly the heavy quark mass $m$ in Eq.~(\ref{eq:HQETaction}) since it can be factored out of correlation functions. The mass renormalizes additively and with $m_{\rm bare}$ we indicate the renormalized mass plus the counterterm.}.

One can devise a procedure to find the first 16 matching conditions by using (in the notation of~\cite{Palombi:2006pu}): different boundary operators ($\{\gamma_5,\gamma_5\}$ or $\{\gamma_k,\gamma_k\}$) in the three point function containing the 4-fermion operator; different boundary-boundary two point functions $f_1^{\rm hl}$, $k_1^{\rm hl}$; different values of the "SF momenta" $\theta$; different insertion times of the 4 fermion operator; different wave functions for the boundary operators.
We plan a one-loop computation in order to find an optimal set of matching conditions.

As shown in~\cite{Palombi:2006pu}, the symmetries of the lattice 
static action simplify the renormalization pattern of dimension-6 operators
in the PO sector. This could be exploited by using tmQCD at maximal twist in order to compute the 
matrix elements of $Q_1'$ and $Q_2'$ trough the PO operators $\cQ_1'$ and $\cQ_2'$ (in the twisted basis), thus avoiding the mixing with $Q_3'$ and $Q_4'$; cf. eq.~(\ref{eq:tmQCD}).
The use of tmQCD for the light quarks in the matching procedure (which is performed in small volume) would require the introduction of the chirally rotated SF~\cite{Sint:2010eh}.

However, the simplification of the mixing pattern does not appear to carry over to the $O(1/m)$ terms since in the tmQCD framework there are 12 dimension-7 operators (i.e. the same number as for standard Wilson fermions) having the same symmetries of $\cQ_1'$ and $\cQ_2'$ which have to be included in the expansion Eq.~(\ref{eq:O_1oM}).

\vspace*{0.4cm}
\noindent {\large \bf Acknowledgements}: We thank R.~Sommer and Y.~Stanev for useful discussions.

\end{document}